# Automatic Recognition of Dyslalia Affecting Pre-Scholars

Pentiuc, Ştefan-Gheorghe
Schipor, Ovidiu-Andrei
Danubianu, Mirela
Schipor, Doina-Maria

1$^{st}$ October 2007

"Stefan cel Mare" University of Suceava
13, str. Universitatii, Suceava, 720229, Romania, tel.: 40230522978
pentiuc@eed.usv.ro, schipor@eed.usv.ro, mdanub@eed.usv.ro

**Abstract**

This article describes the recognition part of a system that will be used for personalized therapy of dyslalia affecting pre scholars. Dyslalia is a speech disorder that affect pronunciation of one ore many sounds. The full system targets interdisciplinary research (computer science, psychology, electronics) - having as main objective the development of methods, models, algorithms, System on Chip architectures with regards to the elaboration and implementation of a complete system addressing the therapy of dyslalia affecting pre scholars, in a personalized and user centered manner.

The system addresses the number of 10% of children with age between 4 and 7 that, according to the statistics, present different variations of speech impairments. Although these impairments do not create major difficulties concerning common communication, it has been noticed that problems are likely to appear affecting negatively the child's personality as well as his social environment.

## 1    Introduction

There is a powerful preoccupation at the European level in helping the people with speech disorders; that is why the EU Quality of Life and Management of Living Resources program has been developed. The OLP (Ortho-Logo-Paedia) project [1] for speech therapy has been started in 2002; the EU finances it and it is a  complex project, involving the Institute for Language and Speech Processing in Athens and seven other partners from the universitary and medical domains. The scope aims to accomplish a three – modules system (OPTACIA, GRIFOS and TELEMACHOS) capable of interactively instructing the children suffering from dysarthria (difficulty in articulating words due to disease of the central nervous system). The proposed interactive environment is a visual one and is adapted to the subjects' age (games, animations). The audio and video interface with the human subject will be the OPTACIA module, the GRIFOS module will make pronunciation recognition and the computer aided instructing will be integrated in the third module – TELEMACHOS.

The priorities on the international level are represented by the developing information systems that will allow the elaboration of personalized therapeutical paths. The following main directions are considered:

development of expert systems that personalizing the therapeutically guides to the child's evolution;

evaluation of the motivation and progresses that the child's achieves.

The most important objectives are determination of methods for the evaluation of speech impairments [2] where the ground truth data set is based on children between 2 years and 2 years and 11 months for the English language. Until now, the only public made results are a few articles about this subject, while the research is in progress.

The potential users of the system are children affected by speech impairments and logopaed professors (speech therapists).

## 2   State of the art

Up to know, the only public available research are the scientifically articles [2] but research is still in progress on the subject. An interesting project is STAR - Speech Training, Assessment, and Remediation [3], started in 2002, but is still in the developing phase. The members (AI. duPont Hospital for Children and The University of Delaware) aim to build a system that would initially recognize phonemes and then sentences. This research group offers a voice generation system (ModelTalker) and other open source applications for audio processing.

On the international level, Speechviewer III developped by IBM [4] that creates an interactive visual model of speech while users practice several speech aspects (e.g. the sound voice or special aspects from current speech). The ICATIANI device developed by TLATOA Speech Processing Group, CENTIA Universidad de las Américas, Puebla Cholula, Pue., México uses sounds and graphics in order to ensure the practice of Spanish mexican pronunciation [5].

Each lesson explains sounds pronunciation using the facial expression with a particular accent on specifying articulation points and the position of the lips. The system includes several animated faces, each of them showing the correct method of vocal pronunciation and providing feedback to the child answers. In this case, if the child's pronunciation matches the system one, the child is rewarded by a smile or warned by a sad face otherwise.

At the national level, little research has been conducted on the therapy of speech impairments, out of which mostly is focused on traditional areas such as voice recognition, voice synthesis and voice authentication. We can mention the studies made in the Psychology and Education Science Department from "Al. I. Cuza" University, Iaşi. These studies have lead to developing software for aided instructing that will provide feedback regarding the oral fluency. Although there are a lot of children with speech disorder, the methods used today in logopaedia are mostly based on individual work with each child. The few existent computer assisted programs in Romania don't provide any feedback. At international level, there are software applications but quite expensive (500$-1500$) and improper for the phonetic specific of Romanian language.

Taking into consideration the fact that Romanian language is a phonetic one that has its own special linguistic particularities, we consider that there is a real need for the development and use of audio-video systems which can be used in the therapy of different pronunciations problems.

From the computer science point of view, the research project that implies the development of an intelligent system capable of doing assisted therapy can be included in a very important research area: Informational technologies in response to society challenges (For health: early diagnosis, personalized therapy).

In conclusion, although present at the world level, the researches are not specific to the Romanian language and at the national level all the researches, excepting those regarding speech disorders, are in incipient phases.

## 3 Objectives and system architecture

The information systems with real time feedback that address pathological speech impairments are relatively recent due firstly by the amount of processing power they require [6]. The progress in computer science allow at the moment for the development of such a system with low risk factors. Children pronunciation will also be used to enrich the existing audio database and to improve the current diagnosis system's performances.

From the recognition perspective, this system must reach some very specific objectives:
- initial and during therapy evaluation of volunteer children and identification of a modality of standardizing their progresses and regresses (at the level of the physiological and behavioral parameters);
- rigorous formalization of an evaluation methodology and development of a pertinent database for the field;
- the development of an expert system for the personalized therapy of speech impairments that will allow for designing a training path for pronunciation, individualized according to the defect category, previous experience and the child's therapy previous evolution;

- the development of a therapeutically guide that will allow mixing classical methods with the adjuvant procedures of the audio visual system and the design of a database that will contain the set of exercises and the results obtained by the child.

The fact that the system have as an objective to treat the pronunciation disorders in Romanian language gives the system another actuality feature. The high degree of complexity of the project results from the high number of different research areas involved: artificial intelligence ( learning expert systems, data mining techniques, pattern recognition), virtual reality, digital signal processing, digital electronic (VLSI), computer architecture (System on Chip, embedded device), psychology ( evaluation procedures, therapeutic guide, experimental design for validation).

The architecture of the integrated therapy system is presented in figure 1.

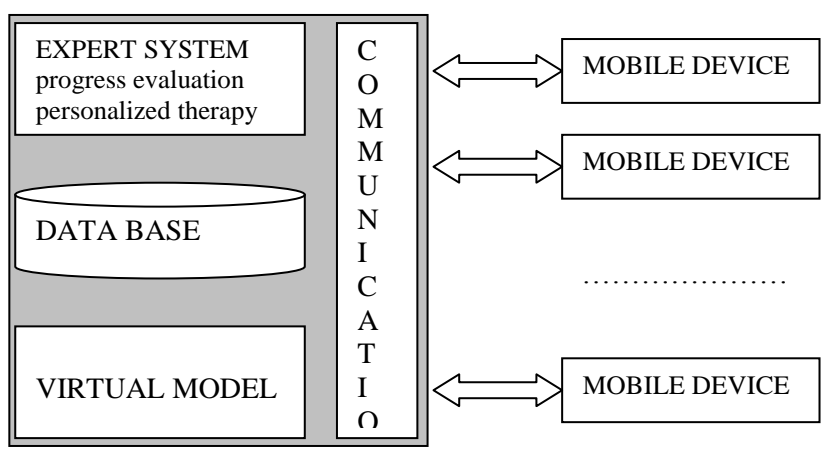

**Figure 1. The architecture of the integrated therapy system**

The functions of the mobile device are as follows:
presetting the child with the speech exercise;
- personalized interaction with the user (based on age, sex, accent differences) during the therapy evaluation of and encouraging the child's progresses;
- capturing audio samples of the child's exercises, communication with the logopaed's PC.

The automatic personalized therapy system stores the prceise evolution and progress of each child and, by adapting the exercises to each child's current level and progress, the speech therapy may take less time to achieve its result. The experimental audio data will be collected and analyzed in the first phase of the research by specialists from the Interschool Regional Logopaedic Center of Suceava. We are also considering the existing research that will provide for an excellent starting point in estimating models that will offer an appropriate visual model for the children.

## 4  The experimental design for system validation

The psychotherapeutic research will follow three steps [7]:
- An experimental stage regarding the application of pre-test – post-test research methodology (through initial and final evaluation) for the comparative analyze between classical and assisted therapeutically ways.
- A diagnostic stage, supposing that we know the attitude of the teachers and the parents regarding the use of a mobile embedded device in dyslalia therapy;
- An integrative stage, regarding the identification of assisted therapy development principles and methods based on previous two stages.

The subjects will be 90 children with pronunciation disorders with the following distribution (3*2 experimental design) [8]:
- Factors: Sigmatism, Rotacism, Polymorph Dyslalia;
- Classical therapy: 15 children, 15 children, 15 children;
- Assisted therapy: 15 children, 15 children, 15 children.

The research will start with a pre-test stage consisting in an initial diagnostic of all children from sample (using specific diagnostic techniques). The same instruments will be applied in post-test stage. Comparative measurements of the evolution will be realized using specific observation scales. This scales will be completed by the logopaed but the results will be offer by the mobile device.

SWOT system analysis (Strengths, Weaknesses, Opportunities, and Threats):
- Strong points: the device will be the result of rigorous research, which will allow changing the normal time spent by a child in front of a computer, with little educational value, to one with raised educational value.
- Weak points: as always when a child uses a computer, some risks are included: the kid could have socialization problems with humans; he could prefer the computer to human contact, stubborn silence if the logopaed does not assist him enough.
- Opportunities: there are possibilities to extend this research to other kind of stubborn silence, possibilities to extend the functionality of the device to other types of stubborn silence and also to other related diseases, the possibility that even some adults will use the device if they have speaking problems, the device functionality can be extended as the client wishes: for a specific age, for a specific speaking problem; also this device can be used also by a child that has parents with no plenty of time to spend with their kids.
- Threats: production cost of mobile device, perceiving the market (T1), averting attitudes from teachers or parents without too many computers knowledge, the logopaeds should learn how to use the mobile devices and computer (T2). Solutions: Threat T1 will be no longer a threat because two factories with high experience in developing of non-conventional devices will be involved in the research project. For T2 a lot adequate dissemination activity will be done.

## 5 System implementation overview

In order to manage children logopaedic activity we design and implement a complex software system named LOGOMON. This system is used by speech therapy teachers for [9]:

- introduction and analyzing of children specific information (automatic obtain special reports);
- making audio recordings with phonemes and score its (for each altered sounds);
- obtain decision support from an integrated expert system;
- creation and evaluation of a large set of exercises for children.

### 5.1 Children information management

LOGOMON is an intelligent system for the individually therapy of the speech disorders. It is a software that performs the specific tasks of a monitor system. These tasks are: the initial evaluation for hunting out the children with speech disorders, their registration in the database, the propose of a possible diagnostic for these children with the possibility for the expert to confirm or to modify this diagnostic, the selection of children for the therapy, the management of the therapy process and the supervision of the children evolution.

Moreover by using LOGOMON we collect and process the data such as we can remove any manual action of collecting or processing of data. Also we can eliminate the data storage on the paper.

The figure 2 presents the main menu of LOGOMON (Administrative Tasks, Complex Examination, Therapy Organizer, Therapy Steps, Reports, Special Operations, About, Exit).

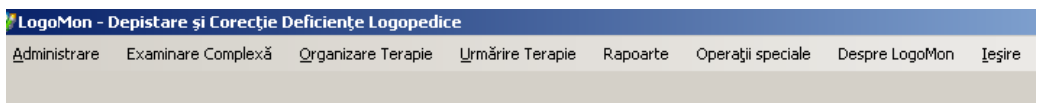

**Figure 2. The main menu of LOGOMON**

### 5.2 Recording and evaluation of children phonemes

In this paper we present our researches regarding automat parsing of audio recordings. These recordings are obtained from children with dyslalia and are necessary for an accurate identification of speech problems. We develop a software application that helps parsing audio, real time, recordings [10].

The main objectives of this task are:
- recording 120 children (60 with correct pronunciation and 60 with dyslalia);
- we must permit different audio environments during recording (some phonemes will be used for training a real time recognition system);
- the cost of recording devices and the children's impact must be minimized;
- after recording is necessary to split the stream into phonemes;
- the speech therapist's voice must be ignored.

We utilize a digital voice recorder in High Quality mode and with VCVA (Variable Control Voice Actuator) activated. The record format is IMA-ADPCM, 16 KHz and 4bits (16 bits PCM). A microphone was placed at 10 cm from mouth in order to minimize environment noise.

A software set of classes (C#) was created for handling audio stream (read, conversion between different format, write).

We also propose an original solution for placing markers in audio stream. These markers are needed for correct parsing of full record.

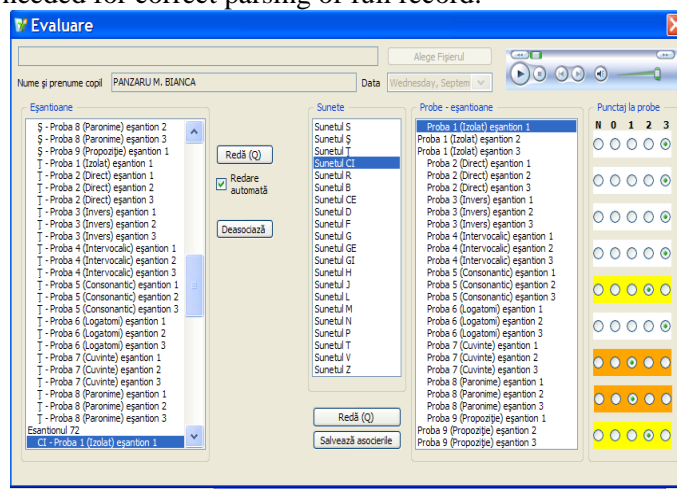

**Figure 3. Recording and evaluation phonemes interface**

In top area, therapist can select and hear original audio file. In left area there is a list with all atomic parts of audio recording (ADM algorithm output). In middle area there are two lists that offer possibility to select expected sound, probe and vocal production. There are some buttons for making/breaking associations. In right area, speech therapist can evaluate probes with scores between 0 and 3. At the end of the process, user can save associations and score for further view/modification.

### 5.3 Expert system

In this project we use a fuzzy expert system for therapy of dyslalic children. With fuzzy approach we can create a better model for speech therapist decisions. A software interface was developed for validation of the system.

The main objectives of this task are:
- personalized therapy (the therapy must be in according with child's problems level, context and possibilities);
- speech therapist assistant (the expert system offer some suggestion regarding what exercises are better for a specific moment and from a specific child);
- (self)teaching (when system's conclusion is different that speech therapist's conclusion the last one must have the knowledge base change possibility).

Fuzzy logic has ability to create accurate models of reality. It's not an "imprecise logic". It's a logic that can manipulate imprecise aspects of reality. In the latest years, many fuzzy expert systems were developed [11].

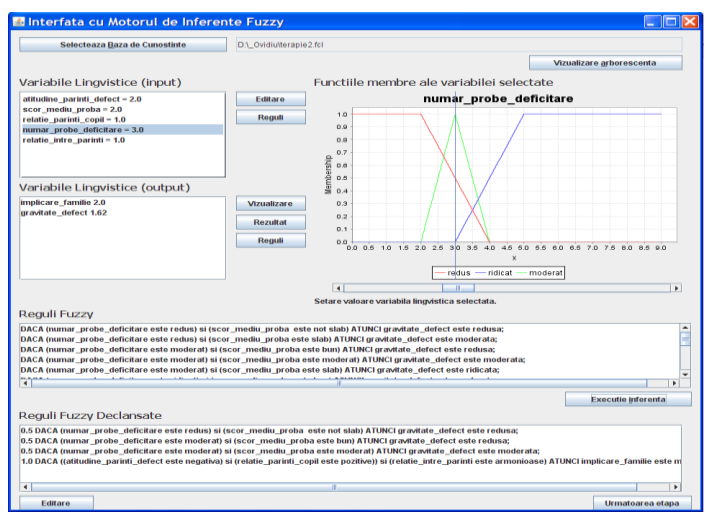

**Figure 4. Expert system validation interface**

For fuzzy rules and variables representation we utilize FCL (Fuzzy Control Language) standard. In order to validate expert system, was developed a software interface (figure 4).

### 5.4 Exercises generator

In order to help children with dyslalia we create a consistent set of software exercises. This set has a unitary software block (data base, programming language, programming philosophy) and a big number of multimedia items for each Romanian language sound (over 5000 audio recordings and over 1000 image).

Speech therapist has possibility to create and save exercises. He also can transmit these exercises to mobile device of children.

## Conclusion

In order to improve speech therapy activity, we develop an integrated system (TERAPERS project). This system is actually tested by Interschool Regional Logopaedic Center of Suceava.